\documentclass[12pt]{iopart}
\newcommand{\BEQ}{\begin{equation}}     
\newcommand{\BEA}{\begin{eqnarray}}
\newcommand{\EEQ}{\end{equation}}       
\newcommand{\EEA}{\end{eqnarray}}
\newcommand{\D}{{\rm d}}
\newcommand{\wht}[1]{\widehat{#1}}      
\newcommand{\gop}{\wht{\phi}_{\vec{0}}} 

\renewcommand{\vec}[1]{{\bf{#1}}}       

\begin{document}

\input epsf.sty



\title{Non-markovian global persistence in phase-ordering kinetics}

\author{Malte Henkel$^1$ and Michel Pleimling$^2$}
\address{$^1$Groupe de Physique Statistique, \\ 
D\'epartement de Physique de la Mati\`ere et des Mat\'eriaux, Institut Jean Lamour\footnote{Laboratoire associ\'e au CNRS UMR 7198},
CNRS -- Nancy Universit\'e -- UPVM, B.P. 70239, \\ F -- 54506 Vand{\oe}uvre l\`es Nancy Cedex, France}
\address{$^2$Department of Physics, Virginia Polytechnic Institute and State University, Blacksburg, Virginia 
24061-0435, USA}
\eads{\mailto{henkel@lpm.u-nancy.fr},\mailto{Michel.Pleimling@vt.edu}}

\begin{abstract}
The persistence probability $P_{\rm g}(t)$ of the global order-parameter of a simple ferromagnet undergoing 
phase-ordering kinetics after a quench from a fully disordered state to below the critical temperature, $T<T_c$, 
is analysed. It is argued that the persistence probability decays algebraically with time in 
the entire low-temperature phase. 
For Markov processes, the associated global persistence exponent 
$\theta_g = \bigl( 2\lambda_C -d\bigr)/(2z)$ is
related to the autocorrelation exponent $\lambda_C$. 
This relationship is confirmed for phase-ordering in the exactly solved $1D$ Ising model and the
$d$-dimensional spherical model. For the $2D$ Glauber-Ising model, the temperature-independent estimate 
$\theta_g=0.063(2)$ indicates that
the dynamics of the global order-parameter is described by a non-Markovian process. 
\end{abstract}
\maketitle

Persistence properties of a stochastic process have been studied
in a variety of dynamical systems, ranging from magnetic systems \cite{persis1,persis2,Stau94,persis3,persis0,persis4,Maj96,Cue96,Cue96b} to soap bubbles
\cite{Tam97}, from reaction-diffusion systems \cite{Odo01} to fluctuating interfaces \cite{Kru97},
turbulent liquid crystals \cite{Take09} 
and non-equilibrium surface growth \cite{Con04}. The persistence
probability $P(t)$ is defined in a very general way as the probability that a given stochastic
variable (observable) retains a characteristic feature over the time $t$. One commonly studied problem
involves the question whether the stochastic variable, which can be a local or a global
quantity, maintains its sign up to the time $t$.

Over the years, the persistence properties of the fluctuations around the mean value of the local order parameter 
$\phi(t)$ have attracted much interest. 
Problems of this kind were initially formulated for the probability $P_{\ell}(t)$ that the
local order-parameter $\phi(t,\vec{r})$ has not yet changed its sign up to time $t$. For large enough times, one
expects a power-law decay $P_{\ell}(t)\sim t^{-\theta_{\ell}}$ if the system is at an equilibrium critical point, 
and where $\theta_{\ell}$ is the {\em local} persistence exponent. This has been studied for simple diffusion 
\cite{Maj96b,Der77} and for one-dimensional 
exactly solvable models quenched to $T=0$ \cite{persis1,persis2,persis3,persis0,persis4}, for example
in the $q$-states Potts model one has the remarkable result
$\theta_{\ell}=-\frac{1}{8}+\bigl[\frac{2}{\pi}\arccos\bigl(\frac{2-q}{\sqrt{2\,} q}\bigr)\bigl]^2/2$ 
\cite{persis0}. Values of $\theta_{\ell}$ for higher-dimensional models have been extracted from numerical
simulations, see \cite{Stau94} for data on the $2D$ Ising model. 

It has turned out to be very useful \cite{Maj96} to 
study the persistence properties of the {\em global} order-parameter
\BEQ
\wht{\phi}_{\vec{0}}(t) =  \frac{1}{\sqrt{|\Omega|\,}} 
\int_{\Omega} \!\D \vec{r}\, \phi(t,\vec{r}) 
\EEQ
averaged over spatial domains $\Omega$ with large volumes 
$1\ll |\Omega|\ll |V|$, where $|V|$ is the total volume of the system. 
The global persistence probability $P_{\rm g}(t)$ is defined as the probability that 
$\gop(t)$ has not changed sign up to time $t$. For large times, one expects a power-law decay 
$P_{\rm g}(t)\sim t^{-\theta_g}$ which defines the {\em global} persistence exponent $\theta_g$. 
This scenario was shown to apply in particular to non-equilibrium critical dynamics ($T=T_c$) \cite{Maj96}. 
Here, an application of the central limit theorem shows that $\gop(t)$ should be described by a Gaussian stochastic 
process and if that process is furthermore Markovian, one  has the scaling relation
$\theta_g z = \lambda_C -d +1 -\eta/2$ \cite{Maj96}. The definition of the 
exponents $\eta$ and $\lambda_C$ will be recalled below. For critical systems, this relation is 
satisfied in a few exactly solvable systems such as the $1D$ Glauber-Ising model quenched to
$T=0$ or the spherical model in $2<d<4$ dimensions (for all of which the dynamical exponent happens to be $z=2$). 
However, the scaling relation does break down in more generic systems, where $z\ne 2$ in all known examples.
One concludes \cite{Maj96} that the process describing the long-time collective dynamics of 
$\gop(t)$ and which arises
from renormalised field-theory, is in general non-Markovian, see \cite{Maj99} for a review. This has been confirmed
by a large variety of analytical and simulational studies, either for critical ferromagnets (including
effects of disorder) \cite{Pau05} or else for genuine non-equilibrium systems at an absorbing phase transition \cite{Oer98}. 
More recent developments consider the cross-over in the global persistence as a function of the initial value
of the magnetisation \cite{Pau07} or the behaviour in the vicinity of surfaces \cite{MajIgl,PleIgl} 
or on sub-manifolds \cite{MajBra}. 

Aging phenomena, which are observed in a broad variety of systems
with slow relaxation dynamics \cite{Bra94,Bou00,God02,Cug02,Hen09}, are paradigmatic examples  
of systems that are far from stationarity, and of great potential for applications.
Initially found in glassy systems, they also occur in
simple ferromagnets, which are often considerably easier to analyse. This kind of
non-equilibrium collective phenomena may be conveniently realized by quenching a system from 
a fully disordered initial state to a temperature $T\leq T_c$, where $T_c$ is the critical
temperature. The specific aspects of aging, namely (i) slow, non-exponential relaxation, (ii)
breaking of time-translation-invariance  and (iii) dynamical scaling, are conveniently studied
through the behaviour of two-time quantities such as the two-time correlator 
$C(t,s;\vec{r})=\langle\phi(t,\vec{r})\phi(s,\vec{0})\rangle$ of the order parameter. For simple ferromagnets,
one finds a dynamical scaling regime when both times are sufficiently large and if 
$t-s\gg t_{\rm micro}$, where $t_{\rm micro}$ is a microscopic reference  time. Then
\BEQ
C(t,s;\vec{r}) = s^{-b} f_C(t/s,|\vec{r}|^z/(t-s)) 
\label{gl:C}
\EEQ
where $b=0$ for $T<T_c$ and $b=(d-2+\eta)/z$ for $T=T_c$, where $\eta$ describes the decay of the critical
equilibrium correlations and $z$ is the dynamical exponent. For $y=t/s$ large, one has
$f_C(y,\vec{0})\sim y^{-\lambda_C/z}$ where $\lambda_C$ is the autocorrelation exponent \cite{Fis88,Hus89}, whose
value depends on whether $T<T_c$ or $T=T_c$.

In this work, we wish to return to an analysis of the properties of the global persistence for simple
ferromagnets which undergo phase-ordering kinetics after a quench from a fully disordered initial state, with 
a vanishing average initial global magnetisation $\langle\gop(0)\rangle=0$, to $T<T_c$ with $T_c>0$.\footnote{For
an ordered initial state, however, one expects $P_{\rm g}(t)\to P_{{\rm g},\infty}(T)$ exponentially fast in $t$ and without 
dynamical scaling. For $T<T_c$, there is no analogue to the cross-over observed in \cite{Pau07} for critical quenches.}
It has been known since a long time that $z=2$ \cite{BraRut,BraRutb} and current theories for the growth laws in phase-ordering kinetics usually start from a local differential equation 
$\partial_t \phi = -\delta{\cal F}/\delta \phi$ where ${\cal F}$ is a local
Landau-Ginzburg functional. Since the temperature $T<T_c$ is thought to be irrelevant, one sets $T=0$ and analyses
the long-time behaviour of the solutions of the above equation, after having averaged over the totally disordered initial state \cite{BraRut,BraRutb,Bra94}. 
Although this set-up is perfectly local in time, the behaviour of the global
persistence probability is non-trivial in general. We find: 
\begin{enumerate}
\item The global order-parameter $\gop(t)$ is described by a Gaussian stochastic process and its
global persistence probability $P_{\rm g}(t)\sim t^{-\theta_g}$ decays algebraically. 
\item If the process for $\gop(t)$ is also Markovian, then $\theta_g=\theta_g^{\rm mark}$, where \cite{Cue96,Cue96b}
\BEQ \label{gl:tg} 
\theta_g^{\rm mark} z = \lambda_C - d/2 \geq 0 
\EEQ
The bound follows from the Yeung-Rao-Desai inequality \cite{Yeu96}. 
\item Eq.~(\ref{gl:tg}) is satisfied for several exactly solvable systems such as the $1D$ Glauber-Ising model
at $T=0$ \cite{Maj96} and the spherical model at $T<T_c$ and for $d>2$. These systems are explicitly Markovian. 
\item While the results stated so far come from essentially standard procedures, we also performed simulations of 
the $2D$ Glauber-Ising model quenched to $T<T_c\approx 2.27$ and find
\BEQ \label{gl:valtg}
\theta_g = \left\{ \begin{array}{ll} 
0.062(2) & \mbox{\rm ~~;~ for $T=1.0$} \\ 
0.065(2) & \mbox{\rm ~~;~ for $T=1.5$}
\end{array} \right. ~.
\EEQ
These results do not agree with eq.~(\ref{gl:tg}). Hence, the process $\gop(t)$ is not Markovian in general. 
This example of a presumably non-integrable system also illustrates that the simple value $z=2$ of the
dynamic exponent, as it holds true in phase-ordering kinetics \cite{BraRut,BraRutb}, 
does not imply the Markov property. 
\end{enumerate}
Generalising the lines of thought developed in \cite{Maj96} to quenches below $T_c$ \cite{Cue96,Cue96b}, 
consider the formation of ordered domains. Shortly after the quench, they should be rather small, with a linear size 
$\xi_0$ at most of the order of the lattice constant. For later times, the firmly established dynamical scaling 
\cite{Bra94} (for a proof of dynamical scaling in $2D$ phase-ordering, see \cite{Are07}) leads to a {\em single} 
relevant time-dependent 
length-scale $L(t)\sim t^{1/z}$ and the dynamics of the system is described by the movement of the domain walls of
independent domains of linear sizes larger than $L(t)$, where the volume $|V|\gg L(t)^d$ of the system must be large 
enough. For any given time $t$, the first two moments of the global order-parameter are finite and read 
$\langle\gop(t)\rangle=0$ and $\langle\gop^2(t)\rangle\sim t^{d/z}$. Since one averages over the configuration space of 
disordered initial states \footnote{If that average is not taken, simple scaling may be replaced by multi-scaling, 
see e.g. \cite{rg}}, 
the central limit theorem is applicable, hence the
stochastic process describing the time-evolution of $\gop(t)$ should be Gaussian, also for $T<T_c$ \cite{Cue96,Cue96b}. 
Again because of dynamical scaling, 
the two-time global correlator is $\langle \gop(t)\gop(s)\rangle = s^{d/z} \hat{f}_C(t/s)$ with $t>s$ and the asymptotic
behaviour $\hat{f}_C(y)\sim y^{(d-\lambda_C)/z}$ as $y\to\infty$. Therefore the normalised global autocorrelator ($t>s$)
\BEQ \label{eq:N}
{N}(t,s) := \frac{\langle \gop(t)\gop(s)\rangle}{\sqrt{\langle \gop^2(t)\rangle\langle \gop^2(s)\rangle\:}}
= \hat{f}_N(t/s)
\EEQ
has the asymptotic form $\hat{f}_N(y)\sim y^{-(\lambda_C-d/2)/z}$ as $y\gg 1$. 
Changing temporal variables according to $T=\ln t$, the process describing
$\gop$ becomes stationary. According to Doob's lemma \cite{Doob42}, in this setting the Markov property is equivalent 
to having the {\em exact} correlator ${N}(t,s) = \bar{N}(T,S) = \exp(-\mu |T-S|)$, with $\mu=(2\lambda_C -d)/(2z)$. 
Then Slepian's formula \cite{Sle62} shows
that the global persistence probability indeed decays with time according to 
$P_{\rm g}(t)\sim t^{-\theta_g^{\rm mark}}$ such that eq.~(\ref{gl:tg}) holds true. 

Not surprisingly, eq.~(\ref{gl:tg}) can be confirmed in some exactly solvable models the dynamics of which are 
explicitly Markovian. As a first example, we consider the $1D$ Glauber-Ising model quenched to $T=0$ from a fully 
disordered state. 
Since $T_c=0$ in $1D$, the global persistence in this model was already considered as an example for a critical quench 
\cite{Maj96}, but we shall see that the same result may also be interpreted in a low-temperature setting. Indeed, from 
the well-known results $\lambda_C=1$ and $z=2$ \cite{God00a}, eq.~(\ref{gl:tg}) gives the Markovian prediction
$\theta_g^{\rm mark}=\frac{1}{4}$. The exact calculation gives the expected power-law decay and 
$\theta_g=\frac{1}{4}$ \cite{Maj96}, as it should be. 

As a second example, we consider the spherical model in $d>2$ dimensions quenched to $T<T_c$. 
The spherical model may be formulated in terms of real-valued spin variables $S_{\vec{r}}$ 
subject to the constraint $\sum_{\vec{r}} S_{\vec{r}}^2 = {\cal N}$,
with the total number $\cal N$ of lattice sites. 
The Hamiltonian describes ferromagnetic nearest-neighbour interactions and reads 
${\cal H}=-\sum_{\vec{(\vec{r},\vec{r}')}} S_{\vec{r}}S_{\vec{r}'}$. 
The non-conserved dynamics is given by a Langevin equation, with a totally disordered initial state. Following the
usual lines of its exact solution \cite{God00b}, the two-time spin-spin correlator at momentum $\vec{q}$ reads
$\wht{C}_{\vec{q}}(t,s) = \wht{C}_{\vec{q}}(s) e^{-\omega(\vec{q})(t-s)} \sqrt{ g(s)/g(t)\,}$ where
\BEA
\wht{C}_{\vec{q}}(t) &=& \frac{\exp(-2\omega(\vec{q})t)}{g(t)} \left[ \wht{C}_{\vec{q}}(0) 
+ 2T \int_0^t \!\D\tau e^{2\omega(\vec{q})\tau}g(\tau)\right]
\nonumber \\
g(t) &=& f(t) + 2T \int_0^t \!\D\tau\, f(t-\tau) g(\tau)
\label{gl:Volt}
\EEA
with $f(t) =\bigl( e^{-4t} I_0(4t)\bigr)^d$ and $I_0$ is a modified Bessel function, while
$\omega(\vec{q})=2\sum_{i=1}^d \bigl(1-\cos q_i\bigr)\simeq q^2$. Then the normalised global correlator becomes
\BEQ
{N}(t,s) = \frac{\wht{C}_{\vec{0}}(t,s)}{\sqrt{\wht{C}_{\vec{0}}(t)\wht{C}_{\vec{0}}(s)\:}} = 
\sqrt{ \frac{\wht{C}_{\vec{0}}(s) g(s)}{\wht{C}_{\vec{0}}(t) g(t)}\:} ~.
\EEQ
For a fully disordered initial state $\wht{C}_{\vec{0}}(0)=1$. 

Indeed, for zero temperature $T=0$, we readily have ${N}(t,s)=1$, 
in agreement with the explicit Markov property in this 
model. In order to show that this result remains true for all $0\leq T<T_c$, we write first
$\wht{C}_{\vec{0}}(t) g(t) = 1 + 2T \int_0^t \!\D\tau\, g(\tau)$ and then solve the Volterra integral equation 
(\ref{gl:Volt}) \cite{God00b,Ebb08}, which leads for large enough times to 
$\int_0^t \!\D\tau\, g(\tau) = g_0 + g_1 t^{1-d/2}$, 
where the explicitly known values of the constants $g_{0,1}$ will not be needed. Hence 
\BEQ
{N}(t,s) = 1 + {\rm O}\bigl(s^{1-d/2}\bigr)
\EEQ
indeed follows an exact power law in $t/s$, 
up to corrections to dynamical scaling, and Slepian's formula \cite{Sle62} gives
$\theta_g=0$. Since the autocorrelation exponent $\lambda_C=d/2$ \cite{God00b}, this agrees with the Markovian 
prediction (\ref{gl:tg}), as expected.\footnote{A similar conclusion holds true for the {\em long-range} spherical
model, with Hamiltonian ${\cal H}=-\sum_{\vec{r}\ne\vec{r}'} J(\vec{r}-\vec{r}') S_{\vec{r}}S_{\vec{r}'}$
and $J(\vec{r})\sim |\vec{r}|^{-d-\sigma}$ with $0<\sigma<2$, quenched to $T<T_c$. For non-conserved dynamics, 
one has $z=\sigma$, $\lambda_C=d/2$
\cite{Baum07}, and a straightforward extension of the above calculation shows that $N(t,s)=1$, up to finite-time
corrections, and that eq.~(\ref{gl:tg}) is satisfied.}

\begin{figure}[bt]
\centerline{\epsfxsize=4.00in\ \epsfbox{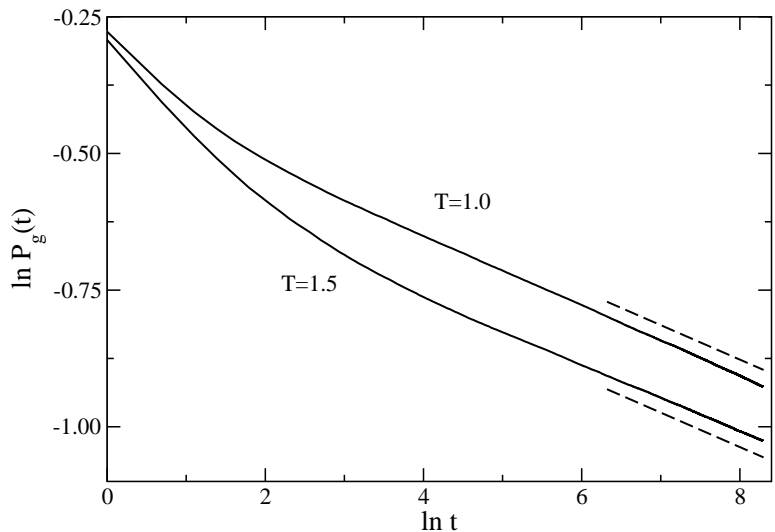}}
\vspace{-2mm}
\caption{Global persistence probability for the $2D$ Glauber-Ising model, quenched either to $T=1.0$ or to $T=1.5$. 
After an initial
regime at short times a power-law regime is observed, with an exponent $\theta_g \approx 0.063$ (dashed lines). Data 
were averaged over $80,000$ different runs with different initial states and different realizations of the noise.
\label{Abb1}
}
\vspace{-4mm}
\end{figure}

Finally, we study the global persistence in the $2D$ Glauber-Ising model, quenched to $T<T_c$ from a fully random
initial state with zero magnetisation. In Fig.~\ref{Abb1} we show the global persistence probability $P_{\rm g}(t)$,
and also compare results for different values of $T$. The global persistence probability is simply the probability that
the global magnetisation has not crossed zero until time $t$. Finite-size effects were monitored carefully by 
simulating systems of various
system sizes. The data in Fig.~\ref{Abb1} have been obtained for a system with $400 \times 400$ sites and result from 
an average over 80,000 different runs where we averaged over different initial conditions  \cite{rg} 
and different realizations of the noise.   
Clearly, for sufficiently large times the decay of the global
persistence is described by a power-law. The values of $\theta_g$ extracted are listed in eq.~(\ref{gl:valtg}). 
However, since $\lambda_C=1.25(2)$ \cite{God02,Hen09}, the estimates (\ref{gl:valtg}) 
are quite distinct from the Markovian prediction 
$\theta_g^{\rm mark}=0.125(2)$. Therefore, 
the effective long-time dynamics of the global order-parameter should {\em not} be described by a Markov process. 
This is our main result. 

The data in Fig.~\ref{Abb1} seem to indicate the existence of an initial short-time regime, 
followed by the power-law decay. In fact,
as we start from a disordered initial state, the dynamical correlation length $L(t)$ increases as a function of time. 
One then
expects that the power-law behaviour characteristic of the $T=0$ fixed point should 
dominate the persistence probability only once $L(t) \gg \xi$
where $\xi$ is the equilibrium correlation length which sufficiently close to 
$T_c$ should scale as $\xi\sim (T_c -T)^{-\nu}$. As long as $L_0 \ll L(t) \ll \xi$, 
where $L_0$ is a microscopic length scale,
the leading behaviour of $P_{\rm g}(t)$ should be the same as at the critical
point. In order to verify this, we monitored the initial time behaviour as a function of temperature and 
found that the initial regime
can be described by a power-law decay with a temperature-dependent {\em effective} exponent. 
The value of this exponent (as well as the cross-over time between the short-time regime and the late-time regime, 
see Fig.~\ref{Abb1}) increases with 
increasing $T$. We find the values 0.18 for $T=1.8$ and 0.20 for $T=2.0$, already close to the value 
$\theta_g=0.237(3)$ \cite{Maj96,Sch97,Pau07} of the global persistence exponent at the critical point.
If $L(t)$ is of the order of $L_0$, the evolution will depend on all sorts of non-universal, microscopic details,
and no general, model-independent statement can be made.

\begin{figure}[t]
\centerline{\epsfxsize=5.50in\ \epsfbox{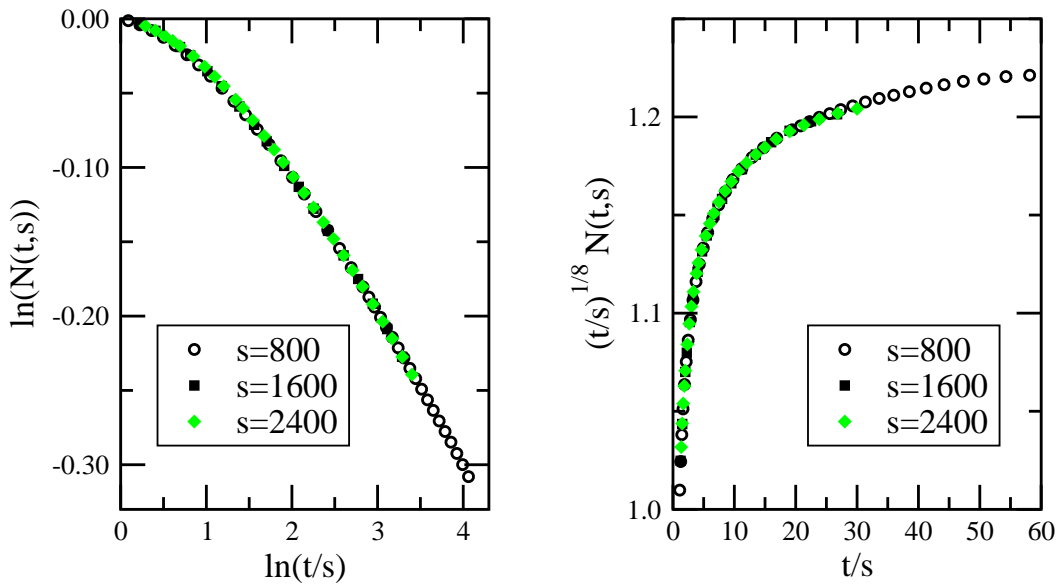}}
\vspace{-2mm}
\caption{(Colour online) Normalised correlator ${N}(t,s)$ for the $2D$ Glauber-Ising model, quenched to $T=1.5$. 
(a) Data obtained for
different waiting times fall on a common master curve when plotted as a function of $t/s$. 
The effective exponent at large
times is $\approx 0.115$, already close to the expected value $0.125$.
(b) Plotting $\left( t/s \right)^{0.125} {N}(t,s)$ over against $t/s$ shows that even for the largest accessible values 
of $t/s$, the asymptotic power-law regime is not yet completely reached.
The error bars, which result from averaging over at least 5,000 independent runs, 
are much smaller than the symbol sizes.
\label{Abb2}
}
\vspace{-4mm}
\end{figure}

In Fig.~\ref{Abb2} we show the normalised correlator ${N}(t,s)$ for different waiting times $s$
at the temperature $T=1.5$. We first verify in
Fig.~\ref{Abb2}a that ${N}(t,s)$ is indeed only a function of $t/s$, see (\ref{eq:N}). As we are exploring the 
time-dependence of
the correlator up to $t/s = 60$, we have to simulate rather large systems with $800 \times 800$ spins in order to
avoid finite-size effects. 
While in the accessed time-window the value of the effective exponent of the 
asymptotic decay $\approx 0.115$ is already close to the expected value $0.125$, 
we also find that this correlator does not
follow a simple power law which gives additional evidence against $\gop$ being described by a Markov process. 

Our result (\ref{gl:valtg}) for $\theta_g$ is consistent with data on the block persistence at $T=0$ \cite{Cue96,Cue96b},  
where $\theta_g\approx 0.09$ in the $2D$ Ising model and $\theta_g\approx 0.06$ in the $2D$ 
time-dependent Ginzburg-Landau
equation was found, in agreement with the expectation that $T$ should be an irrelevant parameter. 

We observe that while for quenches to $T=T_c$, all known systems give $\theta_g \geq \theta_g^{\rm mark}$, see e.g.
\cite{Maj96,Oer98,Maj99,Pau07,MajIgl,PleIgl,Hen09,Sch97}, the
presently available evidence suggests that the opposite inequality $\theta_g \leq \theta_g^{\rm mark}$ might hold true
for quenches into the low-temperature phase $T<T_c$ (here, we exclude cases with $T_c=0$ from the discussion). 

Summarising, we have studied the extension of a scaling relation \cite{Maj96,Cue96,Cue96b} 
for the global persistence exponent $\theta_g$, 
valid for Markovian systems, from critical systems to phase-ordering kinetics. 
We have found evidence, through the breaking of the scaling relation (\ref{gl:tg}) and the
non-trivial form of the normalised global two-time correlator $N(t,s)$, that
the time-dependent global order-parameter $\gop$ should in general {\em not} 
be described by a Markovian stochastic process, unless the underlying model is integrable. 
The example of the $2D$ Ising model quenched to $T<T_c$ illustrates explicitly that 
non-markovian dynamics and a simple value $z=2$ are independent properties of non-equilibrium
critical systems. 
It is tempting to speculate that the generically non-markovian dynamics 
might be related to the fact that in phase-ordering kinetics
the width of the boundaries between the ordered domains increases with time as $w(t)\sim t^{\kappa}$, with
$\kappa=\frac{1}{4}$ in the $2D$ Ising model \cite{Abr89}, 
and which might lead to some `memory effect' in the causal 
interactions of the coarse-grained order-parameter across the interfaces. Testing this further would require 
precise simulational data for a quench right to $T=0$ (when the roughening would be absent) 
and by comparing with our results, obtained for
$T>0$. Further studies of the non-critical 
behaviour of the global persistence $P_{\rm g}(t)$ would be very welcome.  

\ack
We thank S.N. Majumdar and C. Sire for useful correspondence and A. Rosso for a discussion. 
MH thanks the {\it groupe de travail maths/physique} in Nancy for stimulating questions. 
This work was supported in part by the US National
Science Foundation through DMR-0904999.


\end{document}